\documentclass[prl,twocolumn,showpacs,amsmath,amssymb]{revtex4}
\usepackage{epsfig}

\begin{document}

\title{Temperature dependent shifts in collective excitations of a gaseous
  Bose-Einstein condensate}

\author{Juan Jos\'e Garc{\'\i}a-Ripoll}
\affiliation{Departamento de Matem\'aticas, Escuela T\'ecnica Superior de
  Ingenieros Industriales, \\
  Universidad de Castilla-La Mancha 13071 Ciudad Real, Spain}

\date{\today}

\begin{abstract}
  The moment method is applied to the quantum theory for a trapped dilute gas,
  obtaining equations for the evolution of the cloud. These equations proof the
  existence of undamped oscillations in a two-dimensional harmonic trap with
  radial symmetry. In all other cases we can, with physically reasonable
  approximations, solve the equations and find the frequencies of the
  collective modes for all temperatures, from the pure Bose-Einstein condensate
  up to the normal gas.
\end{abstract}

\maketitle

Ever since the first achievement of a weakly interacting Bose-Einstein
condensate in laboratory \cite{anderson,davis,bradley}, one of the basic tools
for its study were the oscillation modes of the condensed cloud
\cite{jin,mewes}. The measure of the frequencies of these oscillations at very
low temperatures, and the comparison with several theoretical predictions
\cite{stringari,prl} were the first important backing to the Gross-Pitaevskii
equation, the mean-field model which describes the condensed cloud in the zero
temperature limit.

Since those first works, there has been a growing interest on the study of the
intermediate regimes, which are far from the zero temperature limit and closer
to the critical temperature $T_c$ at which condensation begins. For $T>0.5T_c$
the collective modes of the three-dimensional condensate suffer both a
frequency shift and a strong damping \cite{jin}. These deviations from the mean
field theory are due to the uncondensed or normal component of the gas. The
role of the the thermal cloud has been partially explained using
self-consistent solutions of the many-body theory for the condensate
\cite{hutchinson,dodd}, theory of relaxation phenomena, linear response theory,
etc [See Ref. \cite{dalfovo} for a list of references].  Nevertheless the
theoretical and experimental studies of these excitations and of their damping
are far from complete, and although the sum of all works gives some insight
into the behavior of the dilute gas, there is no macroscopic theory which
systematically predicts both the finite temperature shifts and the damping of
the excitations.

In this work we attempt a different approach to the study of collective modes.
Our goal is not to consider the condensed and normal components separately, but
to treat the system as a whole. We will start from first principles, studying
the quantum Hamiltonian of the dilute gas and deriving equations for a few
relevant observables: the widths of the cloud, the kinetic energy on each
direction, etc. We will prove that these equations close exactly in the
two-dimensional case with radially symmetric trap. This exact closing reveals
the existence of undamped radial excitations of the cloud, with a universal
frequency $\omega_{2D} = 2\omega$ which is twice the frequency of the trap. In
the last part of this work we will return to the moment equations for arbitrary
geometry and we will close them using a reasonable approximation for the
interaction energy as a function of the widths. With this hypothesis the
dynamical equations can be solved, giving us the temperature dependence of the
excitation frequencies in the cloud as a whole. Finally this dependence is
compared to other works \cite{dodd} and to experiments \cite{jin}.


\textbf{The model.-} The theory for a dilute gas of bosons in a harmonic
confinement is developed up from a simple Hamiltonian
\begin{eqnarray}
  \label{H1}
  \hat{\mathcal{H}} &=& \int \Psi(\mathbf{x})^\dagger
  \left[-\frac{\hbar}{2m}\triangle + V(\mathbf{x},t)\right]
  \Psi(\mathbf{x})d^nx\nonumber\\
  &+&\int \frac{U}{2}\Psi(\mathbf{x})^\dagger\Psi(\mathbf{x})^\dagger
\Psi(\mathbf{x})\Psi(\mathbf{x})d^nx.
\end{eqnarray}
The trapping potential $V(\mathbf{x},t)=
\frac{1}{2}\sum_i\bar\omega_i(t)^2x_i^2$, may have any symmetry, and it may be
even subject to time dependent perturbations.  The coupling constant
$U=4\pi\hbar^2a/m$ measures the interaction among bosons in terms of the s-wave
scattering length $a$ and the contact interaction
$V_{bosons}(\mathbf{x},\mathbf{y})=a\delta(\mathbf{x}-\mathbf{y})$. Finally,
the Hamiltonian has an important hidden parameter which is the dimensionality
of the space, $n\in\{1,2,3\}$, and that has a crucial role on the behavior of
the cloud.

The Hamiltonian $\hat{\mathcal{H}}$ must be understood as an operator on a Fock
space, where the creation and destruction operators obey the commutation rules
\begin{equation}
  \left[\Psi(\mathbf{x}),\Psi^\dagger({\bf y})\right] =
  \delta(\mathbf{x}-{\bf y}),\quad
  \left[\Psi(\mathbf{x}),\Psi({\bf y})\right] = 0.
\end{equation}
Below a critical temperature, $T_c$, the dilute gas experiences a phase
transition which leads to a macroscopic population of the ground state of
$\hat{\mathcal{H}}$.  By neglecting the number of atoms in the thermal cloud
one may approximate $\Psi(\mathbf{x}) \simeq \sqrt{N}\phi(\mathbf{x}) +
\delta{\Psi}({\bf x})$, where $N$ is the total number of particles of the
cloud. The c-number $\phi(\mathbf{x})$ constitutes the order parameter of a
mean-field theory, and it obeys the so-called Gross-Pitaevskii equation
\begin{equation}
  \label{gpe}
  i\partial_t\phi(\mathbf{x},t) =
  \left[-\frac{\hbar}{2m}\triangle + V(\mathbf{x},t) + UN|\phi|^2\right]
  \phi(\mathbf{x},t).
\end{equation}

In the Thomas-Fermi limit for a stationary trap, the kinetic energy of the
wavefunction is neglected with respect to the interaction energy. Then
the system admits stationary solutions $|\phi_{TF}|^2 \propto (\mu -
V({\mathbf{x}}))/UN$ which allows for a self-consistent determination of the
chemical potential in terms of the trapping strength and of the number of
particles, $N$. Using $\phi_{TF}$ one may study the monopolar and quadrupolar
excitations of the condensate \cite{stringari} and compare them to the
experimental results \cite{jin}. The conclusion is that Eq. (\ref{gpe})
provides a good description of the condensate when the thermal cloud is
undetectable, while for higher temperatures the collective modes experience
a strong, temperature dependent shift \cite{jin}. This shift has been sometimes
explained by changing the effective number of particles in the condensate, $N$,
and recomputing the normal modes, or by performing a numerical study of the
Hamiltonian in various self-consistent frameworks \cite{dodd}.

\begin{figure}
{\centering\epsfig{width=0.9\linewidth,file=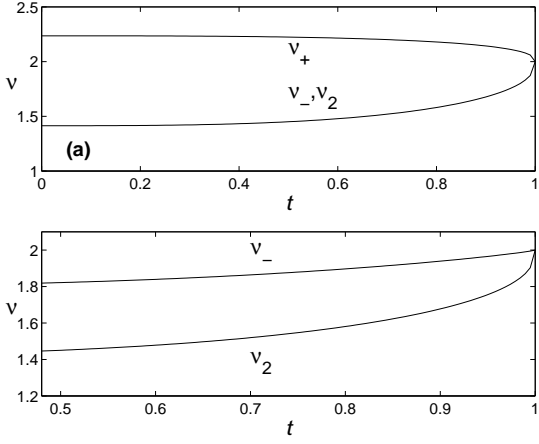}}
\caption{\label{fig-radial}
  Frequencies for monopolar ($\nu_{-,+}$), and quadrupolar ($\nu_2$)
  excitations of a dilute gas, versus normalized temperature $t=T/T_c$. We plot
  the theoretical predictions for (a) spherical trap $\omega_i=1$ and (b)
  pancake trap $\gamma=8$ interpolating the interaction energy between the
  Thomas-Fermi and ideal limits. All figures have been adimensionalized.}
\end{figure}


\textbf{The moment equations.-} The word ``moment'' in this context refers to
the expected value of an observable. If we work with the mean-field theory
these moments are calculated by averaging over the order parameter. For
instance the center of mass of the condensed cloud is
\begin{equation}
  \mathbf{X} = \langle\mathbf{x}\rangle_\phi
  = \int \bar\phi(\mathbf{x})\mathbf{x}\phi(\mathbf{x})d^nx,
\end{equation}
and it follows an equation which is exact, $\ddot{X}_i = - \bar\omega_i(t)^2
X_i$.  These equations and similar ones for other expected values are powerful
tools both in the study of Bose-Einstein condensates and in the study of
nonlinear Schr\"odinger equations in general \cite{torres,thesis}.

We will apply an equivalent technique to the study of the quantum Hamiltonian
(\ref{H}).  First of all we adimensionalize $\hat{\mathcal{H}}$, using as
fundamental units the transverse frequency of the trap,
$\omega_\perp=\sqrt{\omega_x\omega_y}$, and the radial size of this trap, $a_0
= \hbar/\sqrt{m\omega_\perp}$. This way, the Hamiltonian becomes, up to a
global factor,
\begin{eqnarray}
  \label{H}
  \hat{H} &=& \int \Psi(\mathbf{x})^\dagger
  \left[-\textstyle{\frac{1}{2}}\triangle +
    \textstyle{\sum_I\frac{1}{2}\omega_i(t)^2x_i^2}\right]
  \Psi(\mathbf{x})d^nx\nonumber\\
  &+&\int \frac{g}{2}\Psi(\mathbf{x})^\dagger\Psi(\mathbf{x})^\dagger
\Psi(\mathbf{x})\Psi(\mathbf{x})d^nx,
\end{eqnarray}
where $\omega_i=\bar\omega_i/\omega_\perp$ and $g=4\pi a/a_0$. Next we
define a set of one-particle operators
\begin{subequations}
  \label{moments}
\begin{eqnarray}
  \hat{\mathbf{X}} &=&
  \int \Psi^\dagger(\mathbf{x})\mathbf{x}\Psi(\mathbf{x})d^nx,\\
  \hat{\mathbf{P}} &=&
  \int \Psi^\dagger(\mathbf{x})(-i\nabla)\Psi(\mathbf{x})d^nx,\\
  \hat{W}_i &=&
  \int \Psi^\dagger(\mathbf{x})x_i^2\Psi(\mathbf{x})d^nx,\\
  \hat{B}_i &=&
  \int \Psi^\dagger(\mathbf{x})(-ix_i\partial_i)\Psi(\mathbf{x})d^nx,\\
  \hat{K}_i &=& -\frac{1}{2}\int \Psi^\dagger(\mathbf{x})
  \partial^2_i\Psi(\mathbf{x})d^nx,\\
  \hat{J} &=& \int \frac{g}{2}\Psi^\dagger(\mathbf{x})\Psi^\dagger(\mathbf{x})
  \Psi(\mathbf{x})\Psi(\mathbf{x})d^nx,\\
  \hat{F}_i &=& \frac{ig}{2} \int
  \left[(\partial_i\Psi^\dagger)^2\Psi^2-
  (\Psi^\dagger)^2(\partial_i\Psi)^2\right]d^nx.
\end{eqnarray}
\end{subequations}
In these equations $\partial_i$ represents a derivative with respect to $x_i$.
From top to bottom, we have defined the center of mass, its moment, the width
of the cloud, the rate of change of the width, the kinetic energy along the
$i$-th direction, the interaction energy and the rate of change of $J$ along
the $i$-th axis.

We use the Heisenberg picture to study the evolution of these operators. In
this image the density matrix of the system dictates some initial conditions,
$\rho$, and the time evolution is carried by the observables. Thus, any
observable $\hat{A}$ without explicit time dependence follows the simple
equation, $i\frac{d}{dt}\hat{A} = [\hat{A}, \hat{H}]$, and its expected value
is simply $A\equiv\mathrm{Tr}\{A(t)\rho\}$.

The first and most important equation is that of the destruction or creation
operators,
\begin{equation}
  \label{GPE}
  i\partial_t\Psi(\mathbf{x}) = \left[-{\textstyle\frac{1}{2}}\triangle
    +{\textstyle\sum_i\frac{1}{2}}\omega_i(t)^2x_i^2+
    U\Psi^\dagger\Psi\right]\Psi.
\end{equation}
Combining Eqs. (\ref{moments})-(\ref{GPE}), and using the fact that our
operators lack an explicit time dependence, we find the equations for the
center of mass
\begin{equation}
  \label{cm}
  \frac{d^2\hat{\mathbf{X}}}{dt^2} = -\omega_i(t)^2\hat{\mathbf{X}},
\end{equation}
plus a new set of equations for the widths
\begin{subequations}
  \label{full}
  \begin{eqnarray}
    D_t\dot{W}_i &=& B_i,\\
    D_t\hat{B}_i &=& 4K_i - 2\omega_i(t)^2W_i + 2J,\\
    D_t\hat{K}_i &=& -{\textstyle\frac{1}{2}}\omega_i(t)^2B_i - F_i,\\
    D_t\hat{J} &=& {\textstyle\sum_i} F_i,
  \end{eqnarray}
\end{subequations}
where $D_t$ denotes derivative with respect to time.

The first important remark is that Eqs. (\ref{cm})-(\ref{full}) are all exact
and describe the evolution of \emph{one-particle operators}. Nevertheless, one
may now take expected values around all terms in Eqs. (\ref{cm})-(\ref{full})
and the same equations will hold for the expected values
\begin{equation}
\mathbf{X}=\langle{\hat{\mathbf{X}}}\rangle=
\mathrm{Tr}\{\hat{\mathbf{X}}\rho\},
\end{equation} which is on what we
focus from now on. The second remark is that there are less equations than
moments, which seems to prevent us from completely determining the behavior of
the cloud for arbitrary conditions.


\textbf{Undamped oscillations.-} There is an important situation in which the
moment equations may be solved, and that is a two-dimensional condensate in a
radially symmetric trap, $\{n=2,\omega_1=\omega_2=\omega\}$.  In this geometry
we define a total width, $r=\sqrt{W_1+W_2}$, and obtain the following equation
\begin{equation}
  \label{eq2d}
  \frac{d^2r}{dt^2} = -\omega(t)^2r + \frac{2M}{r^3}.
\end{equation}
This equation includes a conserved quantity
\begin{equation}
  M = (K_1 + K_2 + J)(W_1 + W_2) - {\textstyle\frac{1}{8}}(B_1+B_2)^2,
\end{equation}
that must be determined up from the initial conditions.

In the case of an static trap, the equation for the width has an equilibrium
point given by $R=2M/\omega^2$. The oscillations around this point have a
proper frequency, $\omega_{2D}=2\omega$, which is independent of the shape of
the cloud. Therefore, the cloud may be at any temperature and bear excitations
of any multipolarity, but the total width of the cloud will always oscillate
with the universal frequency $\omega_{2D}$ of the monopole mode.

Furthermore, since Eq. (\ref{eq2d}) is exact and it is conservative, any
perturbation of the width of the two-dimensional condensate persist eternally,
and the damping of these oscillations can only be explained in terms of higher
order contributions to $\hat{H}$, such as three- and four-body collisions,
interaction with the environment and anharmonicities of the trap.

One may compare Eq. (\ref{eq2d}) with a similar equation arising from scaling
arguments \cite{kagan}. Those arguments are based on a symmetry of
Eq. (\ref{GPE}) such that it admits infinite many rescaled solutions
\begin{subequations}
\label{scaling2d}
\begin{eqnarray}
  \Psi(\mathbf{x},\tau) &=&
  \frac{1}{\lambda}\Psi_0(\mathbf{x}/\lambda,t)
  e^{-ix^2\dot\lambda/(2\lambda)},\\
  \ddot\lambda &=& -\omega(t)^2\lambda + \lambda^{-3},\quad
  \dot{\tau} = \lambda^2,
\end{eqnarray}
\end{subequations}
for any given solution $\Psi_0({\bf x},t)$. The parameter $\lambda$ can thus be
related to the width of the cloud, but the problem with scalings is that they
implicitly assume a certain type of evolution for the operator, while Eq.
(\ref{eq2d}) makes no such ansatz and it includes further information about
final shape of the collective mode, thanks to $M$.


\textbf{Asymmetric traps.-} We now want to study other dimensionalities and
other geometries of the trap. An inherent limitation in the study of the
moments is that these hierarchies rarely give us closed equations. One must, at
some point make a reasonable approximation to estimate some of the unknowns and
get a tractable model. Roughly, we will assume that the self-interaction energy
is inversely proportional to the effective volume of our $n$-dimensional cloud,
$V=(\Pi_iW_i)^{1/2}$, with a constant $J_0$ that depends on the initial data:
\begin{equation}
  \label{cut}
  J \simeq J_0(\Pi_iW_i)^{-1/2},\quad
  F_i \simeq -{\textstyle\frac{1}{2}}B_iJ/W_i.
\end{equation}

With Eq. (\ref{cut}) we can close the moments hierarchy, obtaining $2n$
differential equations for $2n$ moments
\begin{subequations}
\label{eq3d}
\begin{eqnarray}
  \frac{d^2x_i}{dt^2} &=&
  -\omega_i(t)^2x_i + \frac{2M_i}{x_i^3} + \frac{J}{x_i},\\
  \frac{dM_i}{dt} &=& J \frac{dx_i^2}{dt},
\end{eqnarray}
\end{subequations}
with three new variables $M_i = K_iW_i - \frac{1}{8}B_i^2.$

The motivation for the ansatz (\ref{cut}) is multiple. First it arises from
scaling considerations. If we assume that $\{W_1\ldots W_n\}$ are the only
independent variables of our model, and we rescale our model according to
$x_i\rightarrow \lambda_i x_i$, the destruction operator, the widths and the
interaction change as $\Psi\rightarrow \Psi/\Pi_i\lambda_i$, $W_i\rightarrow
\lambda_i^2 W_i$, and $J\rightarrow J/\Pi_i\lambda_i$, which immediately leads
to Eq. (\ref{cut}).

The other motivation for Eq. (\ref{cut}) is that it works in the mean field
theory too. Making simulations of Eq. (\ref{gpe}) with wavefunctions of almost
any reasonable shape, it is readily seen that $J_0$ is conserved up to a
$10\%$, and even in the worst cases the model equations
(\ref{eq3d}a)-(\ref{eq3d}b) are extremely accurate \cite{thesis}.


\textbf{Collective modes.-} Eqs. (\ref{eq3d}a)-(\ref{eq3d}b) model the
evolution of the widths of a dilute gas at any temperature. They do not make
any distinction between condensed and normal components and the role of
temperature is to change the equilibrium values of $M_i$ and $J$. In order to
study the collective modes of the condensate it is easier to go back to Eqs.
(\ref{full}a)-(\ref{full}d) and linearize around the equilibrium points
$\{W_i,B_i,K_i,J\}$. To first order in the small variables $\{w_i,b_i,k_i\}$,
with the ansatz (\ref{cut}), we get
\begin{subequations}
\label{linearized}
\begin{eqnarray}
  \dot{w}_i &=& b_i,\\
  \dot{b}_i &=& 4k_i - 2\omega_i^2w_i - \sum_j\frac{J}{W_j}w_j,\\
  \dot{k}_i &=& -\frac{1}{2}\omega_i^2b_i + \frac{J}{2W_i}b_i.
\end{eqnarray}
\end{subequations}

There are two important limits in these equations. In the ideal gas limit the
interaction energy can be neglected, $J=0$. This means that each width
decouples from the rest, and that there are three normal modes which oscillate
with frequencies $\nu_i=2\omega_i$. In the the Thomas-Fermi limit we rather
assume $K_i\ll{}J$ and just keep the interaction energy and the trap.  Then the
equilibrium point is given by $J = \omega_i^2W_i$. For a three-dimensional trap
with axial symmetry, $\omega_1=\omega_2=1$, $\omega_3=\gamma$, the excitation
frequencies become
\begin{equation}
  \label{nu-tf}
  \nu_2^2 = 2,\quad
  \nu_\pm^2 = \frac{3}{2}\gamma+2\pm\frac{1}{2}\sqrt{9\gamma^2-16\gamma+16},
\end{equation}
in full agreement with \cite{stringari}. The first frequency corresponds to a
$m=2$ mode in which the $w_x$ and $w_y$ widths oscillate with opposite phases.
The second two frequencies correspond to two $m=0$ modes such that the
transverse shape of the condensate is preserved and $w_x=w_y$. There are also
three Goldstone modes, $\nu=0$, which arise from the scaling symmetry of Eq.
(\ref{full}).

\textbf{Temperature dependent frequencies.-} If we now focus on the
temperature, the ideal gas limit and the Thomas-Fermi limit qualitatively
describe the $T\geq{T_c}$ and $T=0$ limits, respectively. In between these
extremes the amount of condensed and normal clouds vary, inducing changes on
the values of $\{W_i,J\}$ which cause a continuous evolution of the oscillation
frequencies, from $\{2\omega_i\}$ to values which are close to Eq.
(\ref{nu-tf}).

In the radially symmetric case, Eq. (\ref{linearized}) gives us two different
oscillation frequencies
\begin{equation}
  \nu_{-,2}=\omega\sqrt{4-2P},\quad\nu_+=\omega\sqrt{4+2P},
\end{equation}
where $P=J/(\omega R^2)$ depends on the actual shape of the cloud. A crude
estimate for the interaction energy \cite{dalfovo} assigns $P=1$ and $P=0$ for
the $T=0$ and $T=T_c$ limits. In between these limits the interaction energy is
supposed to scale as $P=(N_0/N)^{2/5}=(1-t^3)^{2/5}$, where $N_0$ is the number
of particles in the condensate, and $t=T/T_c$ is the temperature relative to
the phase transition due to the Bose-Einstein condensation \cite{dalfovo}.
This oversimplified model provides the same frequencies as a condensed cloud
with $N_0$ particles and no thermal cloud, and it agrees well with self
consistent calculations in the Popov theory \cite{dodd}.

As an example, in Figure \ref{fig-radial}(a) we plot the oscillation
frequencies as a function of the relative temperature, $t$, for a spherically
symmetric trap. Better estimates of the excitation frequencies require better
estimates of the functions $\{J(T),W_i(T)\}$. That task is significantly
simpler than full calculations of the excitation spectrum for $N$ particles,
but it is yet an open problem.

Finally we have compared our predictions with the experimental results of D.~S.
Jin \emph{et al.} \cite{jin}. The only difference is on the quadrupolar
excitations, $\nu_2(t)$, whose experimental frequency seems to decrease when
the temperature increases. This point was already notice in \cite{dodd} but now
it becomes evident that the deviation could be due to studying the width of the
condensate and of the thermal cloud separately.  We believe that this apparent
contradiction deserves further experimental investigation, preferably by
studying the oscillations in clouds with more atoms and various geometries.


\textbf{Summary and discussions.-} We have studied the collective modes of a
trapped dilute gas. By computing the evolution of several operators, we have
obtained differential equations for the center of mass and the widths of the
cloud. In the two-dimensional case with radially symmetric trap we have
predicted the existence of undamped oscillations of the radial width with
frequency $\omega_{2D}=2\omega$ twice the frequency of the trap. This
prediction can be confirmed in current experiments \cite{Ketterle}.

For other symmetries and dimensionalities we recovered both the Thomas-Fermi
limit and the ideal gas limit. We also combined a thermodynamical estimate of
the interaction energy with our dynamical equations to obtain the temperature
dependence of the collective modes in various traps, with similar results as
previous numerical works \cite{dodd} and experiments \cite{jin}.

This work offers new techniques that can be applied to the study of trapped
dilute gases in general and to Bose-Einstein condensation in particular. We
suggested a new way to perform experiments, by studying the gas cloud as a
whole, and we also offered a simple way to relate the normal modes of the cloud
to the interaction energy of the gas, without the need to perform difficult
self-consistent calculations.  Finally, with our analysis it becomes clear why
simple models such as scalings \cite{kagan} have been so useful in analyzing
current experiments.


\end{document}